\documentclass[twocolumn,aps,prb,amssymb,superscriptaddress,showpacs]{revtex4-1}

\usepackage{amsmath}
\usepackage{amssymb}
\usepackage{amsfonts}
\usepackage{dsfont}
\usepackage{graphicx}
\usepackage{bm}
\usepackage{appendix}
\usepackage{epsfig}
\usepackage{appendix}

\newcommand{\be}{\begin{equation}}
\newcommand{\ee}{\end{equation}}
\newcommand{\nmajorana}{n}

\usepackage[usenames,dvipsnames]{color}

\begin{document}
\title{Transport signatures of interacting fermions in quasi-one-dimensional topological superconductors}

\author{Dganit Meidan}
\affiliation{\mbox{Department of Physics, Ben-Gurion University of the Negev, BeÕer-Sheva 84105, Israel}}
\author{Alessandro Romito}
\affiliation{\mbox{Dahlem Center for Complex Quantum Systems and Fachbereich Physik, Freie Universit\"at Berlin, 14195 Berlin, Germany}}
\author{Piet W. Brouwer}
\affiliation{\mbox{Dahlem Center for Complex Quantum Systems and Fachbereich Physik, Freie Universit\"at Berlin, 14195 Berlin, Germany}}

\begin{abstract}
A topological superconducting wire with an effective time reversal symmetry is known to have a $\mathbb{Z}_8$ topological classification in the presence of interactions. The topological index $|\nmajorana| \le 4$ counts the number of Majorana end states, negative $\nmajorana$ corresponding to end states that are odd  under time reversal. If such a wire is weakly coupled to a normal-metal lead, interactions induce a Kondo-like correlated state if $|\nmajorana| = 4$. We show that the Kondo-like state manifests itself in an anomalous temperature dependence of the zero-bias conductance and by an anomalous Fano factor for the zero-temperature normally-reflected current at finite bias. We also consider the splitting of the effective Kondo resonance for weak symmetry-breaking perturbations.
\end{abstract}

\maketitle

\section{Introduction} 

Topological superducting wires in one dimension differ from their non-topological counterparts by the presence of a pair of zero-energy Majorana bound states at the two ends of the wire.\cite{Kitaev2001} Multiple Majorana bound states at each end  do not occur generically, since a pair of Majorana bound states combines to a standard (Dirac) fermion, with an energy that is no longer pinned to zero. Additional symmetries may, however, prevent multiple Majoranas from acquiring a finite energy and, hence, protect a topological phase with more than one Majorana bound state at the wire's ends. In an effectively spinless system, time-reversal symmetry permits the coexistence of an arbitrary number of Majoranas at the same end of the wire, as long as these have the same parity under time-reversal. In a more formal language, for a spinless system the presence of time-reversal symmetry changes the topological classification from $\mathbb{Z}_2$ ({\em i.e.}, the numbers $0$ or $1$) to $\mathbb{Z}$,\cite{Schnyder2008,Ryu2010,Kitaev2009} where the positive/negative numbers refer to Majorana states that are even/odd under time reversal.

Experimentally, topological superconducting wires can in principle be engineered in hybrid structures involving proximity-induced superconductivity in topological insulators,\cite{Fu2008} or semiconducting\cite{Lutchyn2010,Oreg2010} and magnetic nanostructures.\cite{Duckheim2011,Chung2011,Kjaergaard2012,Choy2011,Martin2012,Nadj2013} Recent tunneling spectroscopy experiments indeed reported possible signatures of Majorana bound states in semiconductor wires with a strong spin-orbit coupling, proximity coupled to a superconductor,~\cite{Mourik2012,Das2012,Higginbotham2015} and in a chain of magnetic atoms on a superconducting substrate.\cite{Nadj2014} 

Although time-reversal symmetry is broken in all these realizations by an applied magnetic field or by the presence of magnetic structures, an effective time-reversal symmetry is predicted to be present as a excellent {\em approximate} symmetry.\cite{Tewari2011,GKellsPRB2012a} This effective time-reversal operation ${\cal T}$ squares to one, {\em i.e.}, it is of the ``spinless'' type, which means that in principle such superconducting wires can host multiple Majorana bound states at each end. In tunneling spectroscopy experiments the presence of multiple Majoranas shows up as a zero-bias conductance peak of height $2 |\nmajorana| e^2/h$, where $\nmajorana$ is the number of Majorana bound states (where a Majorana that is odd under time reversal is counted as $-1$). The (still outstanding) observation of a conductance peak of quantized value $2 e^2/h$ counts as strong evidence for a Majorana state.

In a seminal work, Fidkowski and Kitaev studied the stability of the topological spinless time-reversal symmetric superconducting phases in the presence of weak local interactions.\cite{Fidkowski2010} They found that in the presence of interactions only eight distinct phases remain:\cite{Turner2011,Fidkowski2011a,Gurarie2011,Manmana2012} Interactions completely lift the ground state degeneracy associated with the presence of $\nmajorana=8$ Majoranas, so that the system becomes topologically trivial. Alternatively, with interactions one may continuously deform a system with four Majorana end states that are even under time reversal ($\nmajorana = 4$) into a superconductor with four Majorana end states that are odd under time reversal ($\nmajorana = -4$). Since $|\nmajorana| < 4$ Majorana bound states at one end correspond to less than two fermions, the effective low-energy theory for the interacting system with $|\nmajorana|<4$ is essentially equal to that for the non-interacting system. For $|\nmajorana| = 4$ interactions lower the $4$-fold degeneracy of the ground state of the non-interacting theory (counted at each end) to a two-fold degeneracy.

In a recent publication, we have investigated the interacting phases for a spinless superconducting wire with (effective) time-reversal symmetry weakly coupled to a normal-metal lead, which is the geometry relevant for tunneling spectroscopy experiments.\cite{Meidan2014} Our main findings were that, while for $|\nmajorana| < 4$ the zero-bias transport properties were unaffected by interactions, for $|\nmajorana| = 4$ the interacting system forms a Kondo resonance at low temperature. (Properties of phases with $|\nmajorana| > 4$ follow from the fact that with interactions systems with $\nmajorana$ and $\nmajorana - 8$ are topologically equivalent.) For generic parameters, zero-temperature transport through this $|\nmajorana|=4$ Kondo resonance is described by a Fermi-liquid fixed point, which gives precisely the same zero-bias conductance $G = 8 e^2/h$ as in the non-interacting case.\cite{Meidan2014}

In the present article, we investigate the transport of the $|\nmajorana|=4$ topological phases at finite temperature and/or finite bias. In the limit of low (but finite) temperature and voltage, we find three differences between the interacting and non-interacting cases: (i) With interactions, the characteristic energy scale for the dependence on temperture $T$, voltage $V$, and (effective) magnetic field $B$ is the ``Kondo temperature'' $T_{\rm K} \sim \sqrt{W \Gamma} e^{-W/\Gamma}$, where $W$ is the interaction strength and $\Gamma$ the level broadening due to the coupling to the normal-metal contact, whereas this energy scale is $\Gamma$ without interactions. (The effective magnetic field $B$ measures the strength of a perturbation that breaks the effective time-reversal symmetry ${\cal T}$.) (ii) Without interactions, the zero-bias conductance $G(V=0,T)$ is a monotonously decreasing function of temperature, while the temperature dependence of $G(V=0,T)$ is non-monotonous in the presence of interactions. (iii) The Fano factor for the interacting system (defined with respect to the normally-reflected current) is $F' = 10/3$, which should be compared with the value $F' = 2$ for the non-interacting system. These differences can serve as an experimental signature to detect the formation of the topological Kondo phase at the end of a multi-channel Majorana wire.

Below, in Sec.\ \ref{sec:model} we discuss the details of the model we consider. In Sec.\ \ref{sec:3} we review the zero-temperature transport properties for all values of the topological index $\nmajorana$. The remaining Sections specialize to the case $\nmajorana = 4\ (\mod 8)$; other values of $\nmajorana$ are effectively described by the non-interacting theory. The high and low temperature limits $T \gg T_{\rm K}$ and $T \ll T_{\rm K}$ are discussed separately, in Secs.\ \ref{sec:4} and \ref{sec:5} respectively. In Sec.\ \ref{sec:6} we consider weak perturbations that break the effective time-reversal symmetry responsible for the protection of the $\nmajorana$ Majorana end states in the non-interacting system. We conclude in Sec.\ \ref{sec:conclusions}.

\section{Model}
\label{sec:model}

\begin{figure}

\includegraphics[width=0.45\textwidth]{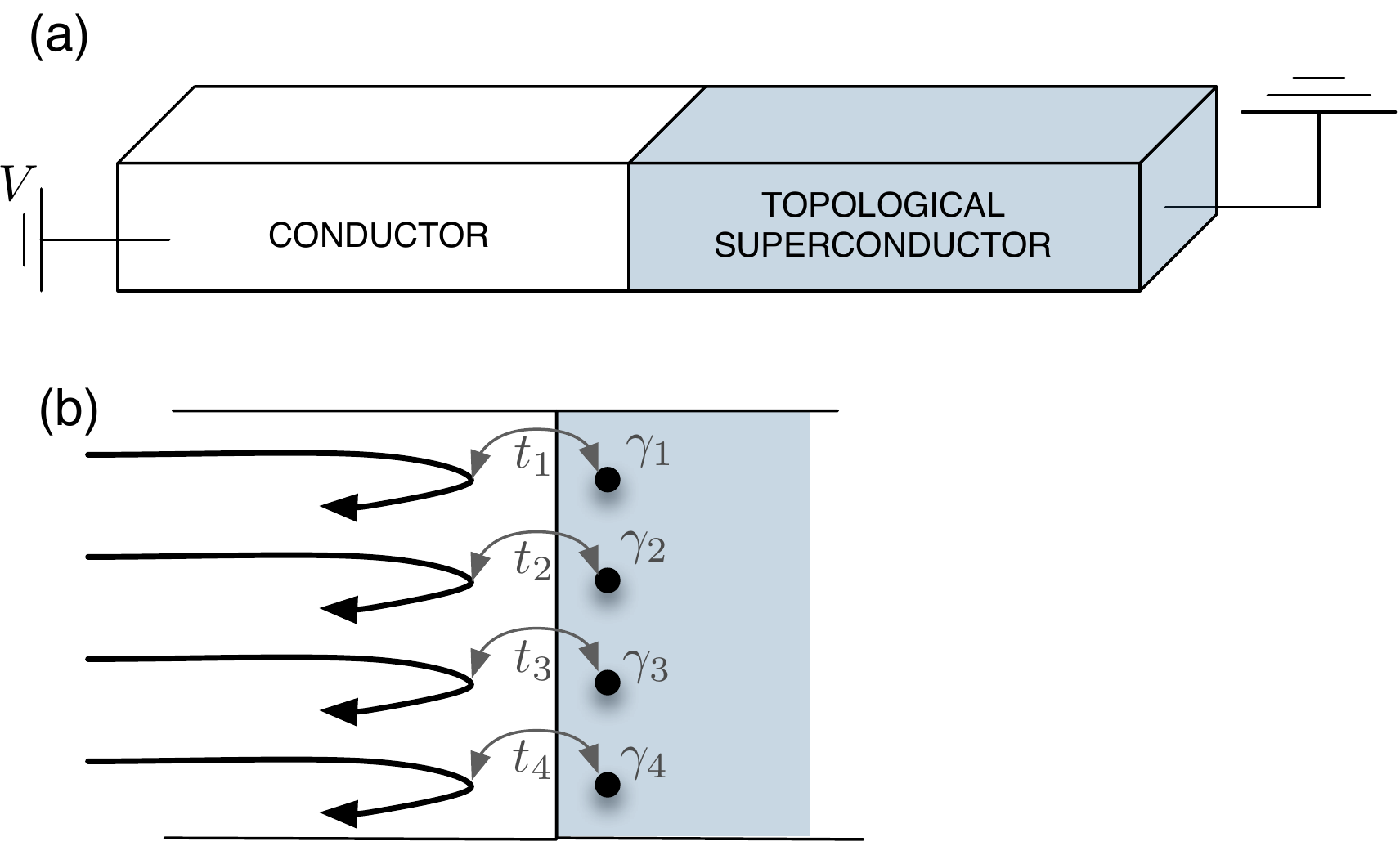}
\caption{\label{fig1}%
(a) Sketch of the system. A voltage-biased quasi-one-dimensional conductor is in
contact with a topological superconductor. Both the conductor and the
superconductor have multiple transverse modes. 
(b) Schematic picture of the model for the $n=4 $ channel case. The
superconductor is described by its low-energy Majorana end-states 
$\gamma_j$, which are tunnel coupled to the scattering states of the lead.}
\end{figure}

The system we consider is shown schematically in Fig.\ \ref{fig1}. It consists of a half-infinite ideal lead, coupled to a half-infinite multichannel topological superconducting wire with an antiunitary symmetry ${\cal T}$ squaring to unity, corresponding to symmetry class BDI of the Cartan classification.\cite{Altland1997} There are short-range two-fermion interactions in the Majorana wire; the electrons in the ideal lead are assumed to be non-interacting.  Following common practice in the field, we refer to the antiunitary effective symmetry ${\cal T}$ as ``time-reversal symmetry'', although one should keep in mind that the true time-reversal symmetry is broken in all of the proposals to physically realize a topological superconducting wire in symmetry class BDI. The effective time-reversal symmetry is a good approximate symmetry in generic topological superconductors with a wire geometry.\cite{Tewari2011,GKellsPRB2012a} In semiconductor-based platforms the existence of multiple channels require a Zeeman field strong enough to induce superconductivity in more than one transverse mode; in ferromagnet-based platforms, effective exchange field are much larger and multiple channels appear naturally in a quasi-one-dimensional geometry.\cite{Duckheim2011}

The ideal lead is described by the Hamiltonian
\be
  H_{\rm lead} = H=\int d\xi \sum_\alpha \xi c^\dagger_{\xi,\alpha} c_{\xi,\alpha},
  \label{eq:Hlead}
\ee
where $\xi$ is the energy, measured with respect to the Fermi level, $\alpha$ is the channel index, and the operators $c_{\xi,\alpha}$ and $c^{\dagger}_{\xi,\alpha}$ annihilate and create a scattering state in channel $\alpha$ at energy $\xi$. The scattering states are assumed to be even under the effective time-reversal symmetry,
\be
  {\cal T} c_{\xi,\alpha} {\cal T}^{-1} = c_{\xi,\alpha}.
\ee

The low-energy degrees of freedom of the topological superconductor wire are encoded in up to four Majorana operators $\gamma_{\alpha}$ localized near the wire's boundary. (A larger number of Majoranas is unstable against local interactions,\cite{Fidkowski2010} so that it need not be considered separately.) Microscopically, the Majorana operators can appear as zero-energy boundary modes of a Kitaev chain,\cite{Kitaev2001} or of a continuum Hamiltonian for a spinless p-wave superconductor. The presence of the effective time-reversal symmetry requires the order pararameter to be real, whereby the sign of the order parameter determines whether the Majorana operators are even or odd under the effective time-reversal operation.\cite{Note1} We denote the number of Majorana operators with $\nmajorana$, with the convention that we choose $\nmajorana$ positive if the Majoranas are even under time reversal,
\be
  {\cal T} \gamma_{\alpha} {\cal T}^{-1} = \gamma_{\alpha},
\ee
and negative if the Majoranas are odd under time reversal. 

The coupling between the normal lead and the topological superconductor is described by a tunneling Hamiltonian, which, after suitable basis changes for the scattering channels in the ideal normal-metal lead and for the Majorana end states, can be made diagronal in the channel index $\alpha$, 
\be
 H_T= \frac{1}{2} \sum_{\alpha=1}^{|n|}  \int d \xi t_\alpha (c^\dagger_{\xi,\alpha} -c_{\xi,\alpha}) \gamma_\alpha,
  \label{eq:HT}
\ee
where our normalization of the fermion creation and annihilation operators implies that the tunneling rate
\be
  \Gamma_{\alpha} = 2\pi|t_{\alpha}|^2.
\ee
We have assumed that the number of channels in the ideal lead
precisely matches the number $|\nmajorana|$ of Majorana bound states
in the superconductor.\cite{Beri2012, Altland2013,Note2}

Finally, upon projecting down to the low-energy Hilbert space, short-range two-fermion interactions exist only for the case of four Majoranas, in which case one has the interaction Hamiltonian
\be
  H_{\rm int} = -W \gamma_1 \gamma_2 \gamma_3 \gamma_4.
  \label{eq:Hint}
\ee

\section{Zero-temperature transport} 
\label{sec:3}

In the absence of interactions, the zero-bias, zero-temperature conductance $G$ of the system is
\be
  G = \frac{2 e^2}{h} |\nmajorana|. \label{eq:Gquantized}
\ee
This result is well known for the case $|\nmajorana|=1$ of a single Majorana.\cite{Flensberg2010,Law2009} Its generalization to arbitrary $\nmajorana$ follows from the observation that $\nmajorana = \mbox{tr}\, r_{\rm eh}$,\cite{Fulga2011} where $r_{\rm eh}$ is the reflection matrix for Andreev scattering at the normal-metal--superconductor interface, together with the well-known expression $G = (2 e^2/h) \mbox{tr}\, r_{\rm eh} r_{\rm eh}^{\dagger}$ for the conductance $G$ of a normal-metal--superconductor junction.\cite{BTK1982,Lambert1991,Takane1992} The conductance (\ref{eq:Gquantized}) is independent of the value of the tunnel amplitudes $t_{\alpha}$ of Eq.\ (\ref{eq:HT}).\cite{Note3} 

At zero temperature, the inclusion of interactions was found to have remarkably little effect on the zero-bias conductance.\cite{Meidan2014} This observation follows immediately for $|\nmajorana|<4$, because in that case interactions have zero matrix elements in the low-energy sector of the Hilbert space. (It takes at least four Majorana operators for an interaction Hamiltonian.) It also follows for $|\nmajorana| > 4$, but $\nmajorana$ not an odd multiple of $4$, because with interactions $\nmajorana$ values that differ by a multiple of $8$ are topologically equivalent,\cite{Fidkowski2010} and a generic perturbation of the Hamiltonian drives the system to a state in which $|\nmajorana|$ takes its minimal value in the topological equivalence class.\cite{Meidan2014} As in the non-interacting case, this conclusion relies on the limit that the tunneling amplitudes $|t_{\alpha}| \downarrow 0$, to ensure that ``gapped-out'' Majorana states do not contribute to the conductance.

Finally a Majorana wire with $|\nmajorana|=4$ Majorana end states is characterized by emergent many-body end states, which show a Kondo-like resonance at zero temperature.\cite{Meidan2014} For generic model parameters this Kondo-like resonance is described by a Fermi liquid fixed point with a conductance $G = 8 e^2/h$. The same conclusion applies to larger $\nmajorana$ if $\nmajorana$ is an odd multiple of $4$.

In this article we will also consider a second transport property, the shot noise power, which is characterized by the Fano factor $F = S/2e|I|$, the ratio of zero-bias shot noise power $S$ to the zero-bias electrical current $I$. In terms of the Andreev reflection matrix $r_{\rm eh}$, the Fano factor for a normal-metal--superconductor junction reads\cite{Jong1994}
\be
  F = 2 \frac{\mbox{tr}\, r_{\rm eh} r_{\rm eh}^{\dagger} (1 - r_{\rm eh} r_{\rm eh}^{\dagger})}{\mbox{tr}\, r_{\rm eh} r_{\rm eh}^{\dagger}}.
\ee
In spite of the presence of interactions, at zero temperature, this equality can be applied to the present system, because an effective Fermi-Liquid description applies to all values of the Majorana number $\nmajorana$. Since all eigenvalues of $r_{\rm eh} r_{\rm eh}^{\dagger}$ are either zero or one at zero temperature, we conclude that $F = 0$ at zero temperature.

In the next two sections we investigate transport away from the zero temperature limit. Section \ref{sec:4} addresses the conductance $G$ in the limit of high temperatures (but $T$ still small enough that the minimal description in terms of $|\nmajorana|$ Majorana states remains valid), while Sec.\ \ref{sec:5} addresses the leading finite-$T$ corrections to the zero temperature results for conductance and shot noise listed here.

\section{High-temperature limit} \label{sec:4}

At high temperature the effect of interactions in the Majorana wire can be treated perturbatively, which allows to obtain an analytic expression for the transport properties. Interaction effects are relevant for $|\nmajorana|=4$ only. In that case the condition for the validity of the perturbation theory is that the temperature $T$ be large in comparison to the Kondo temperature $T_{\rm K} \sim \sqrt{W \Gamma} e^{-W/\Gamma}$ for the equivalent Kondo problem, see Ref.\ \onlinecite{Meidan2014} and Sec.\ \ref{sec:5}. For $|\nmajorana| < 4$ interaction effects are absent, and the results derived below continue to be valid down to zero temperature.

To calculate the current $I$ we start from the expression $I = -e \dot N$, where $N = \int d\xi \sum_{\alpha} c_{\xi,\alpha}^{\dagger} c_{\xi,\alpha}$ is the number of electrons in the ideal lead. From the Heisenberg equation of motion one then obtains,
\begin{eqnarray}
  I &=& \frac{i e}{\hbar} [N,H_{\rm T}] \nonumber \\
  &=& \frac{i e}{2 \hbar} \int d\xi \sum_{\alpha=1}^{|\nmajorana|}  t_\alpha
  (c_{\xi,\alpha} + c_{\xi,\alpha}^{\dagger}) \gamma_{\alpha}.
  \label{eq:current}
\end{eqnarray}
Taking the expectation value and following a well-established procedure,\cite{Meir1992} the current can be expressed in terms of the lesser and greater Green functions
\begin{equation}
   G^{< (>)}_{\alpha,\beta} (\omega) \equiv \int
dt e^{i\omega t} G^{< (>)}_{\alpha,\beta} (t)
\end{equation}
of the Majorana states,
\begin{equation}
  G^<_{\alpha,\beta} (t) = i \langle \gamma_\beta (0) \gamma_\alpha (t) \rangle,  G^>_{\alpha,\beta} (t) = -i \langle \gamma_\alpha (t) \gamma_\beta (0) \rangle.
\end{equation}
One finds
\begin{align}
\nonumber
I=&\frac{ie}{4 \hbar}\sum_{\alpha,\beta} \int d\xi
  t_{\alpha} t_{\beta}
  \\ & \mbox{} \times
\left[ G^<_{\alpha,\beta} (\xi)
  +f(\xi-eV) \left( G^>_{\alpha,\beta}
    (\xi)-G^<_{\alpha,\beta} (\xi) \right) \right],
\label{eq:ht}
\end{align}
with $f(\xi) = 1/(1 + e^{\xi/T})$ the Fermi-Dirac distribution function and $V$ the voltage applied to the normal-metal lead. Using the fact that the Majorana Green functions have the properties $G^{<(>)}_{\alpha,\beta}(t) =\delta_{\alpha,\beta} G^{<(>)}_{\alpha,\alpha}(t)$ (see Appendix~\ref{appendix2}), $G_{\alpha,\alpha}^<(-\xi) = - G_{\alpha,\alpha}^>(\xi) $, and $G_{\alpha,\alpha}^>(\xi)-G_{\alpha,\alpha}^<(\xi)= i \textrm{Im} G^{(r)}_{\alpha,\alpha}(\xi)$, one finds that the conductance $G = \partial I/\partial V$ is given by
\begin{equation}
  \label{eq:GThigh}
  G=\frac{e^2}{h} \sum_\alpha \Gamma_\alpha \int d\xi 
  \textrm{Im}\,
  G^{\rm R}_{\alpha,\alpha} (\xi) \partial_\xi f(\xi-eV),
\end{equation}
with $G^{\rm R}_{\alpha,\alpha} (\xi)$ the (Fourier transformed) retarded Green function, 
\be
  G^{\rm R}_{\alpha,\alpha}(t) = -i\theta(t)\langle\{ \gamma_{\alpha}(t),\gamma_{\beta}(0) \} \rangle.
\ee

Up to this point the analysis is exact. The calculation of the retarded Green's functions for the interacting Majorana end-states involves the approximation that correlations between electron dynamics in the dot and
in the lead can be neglected. This is expected to be a valid approximations at
temperatures $T \gg T_{\rm K}$, where the thermal fluctuations in the lead
prevent the development of Kondo like correlations, and has proven to
be a valid approximation for the Kondo problem.\cite{Meir1991}   
The calculations are reported in appendix \ref{appendix2}, where we find the result
\begin{eqnarray}
  G^{\rm R}_{\alpha,\alpha}(\omega) & = & \frac{2}{\omega+i \Gamma_{\alpha}-\frac{4W^2}{\omega+i (\Gamma - \Gamma_{\alpha})}},
  \label{eq:GR}
\end{eqnarray}
with 
\be
  \Gamma = \sum_{\alpha} \Gamma_{\alpha}.
\ee
These expressions differ significantly from those obtained for the
standard Kondo model in Ref.~\onlinecite{Meir1991}, and a few remarks are in order. First, the sign of the interaction $W$ in Eq.\ (\ref{eq:Hint}) is immaterial, since it changes upon a mere relabeling of the channels $\alpha$. This explains the absence of any terms linear in the interaction parameter $W$. Second, as the expressions above correspond to the Green functions in the Majorana basis, they are manifestly particle hole symmetric, which explains why  there is no dependence on the occupation of any local femionic degrees of freedom.

\begin{figure}
\includegraphics[width=0.45\textwidth]{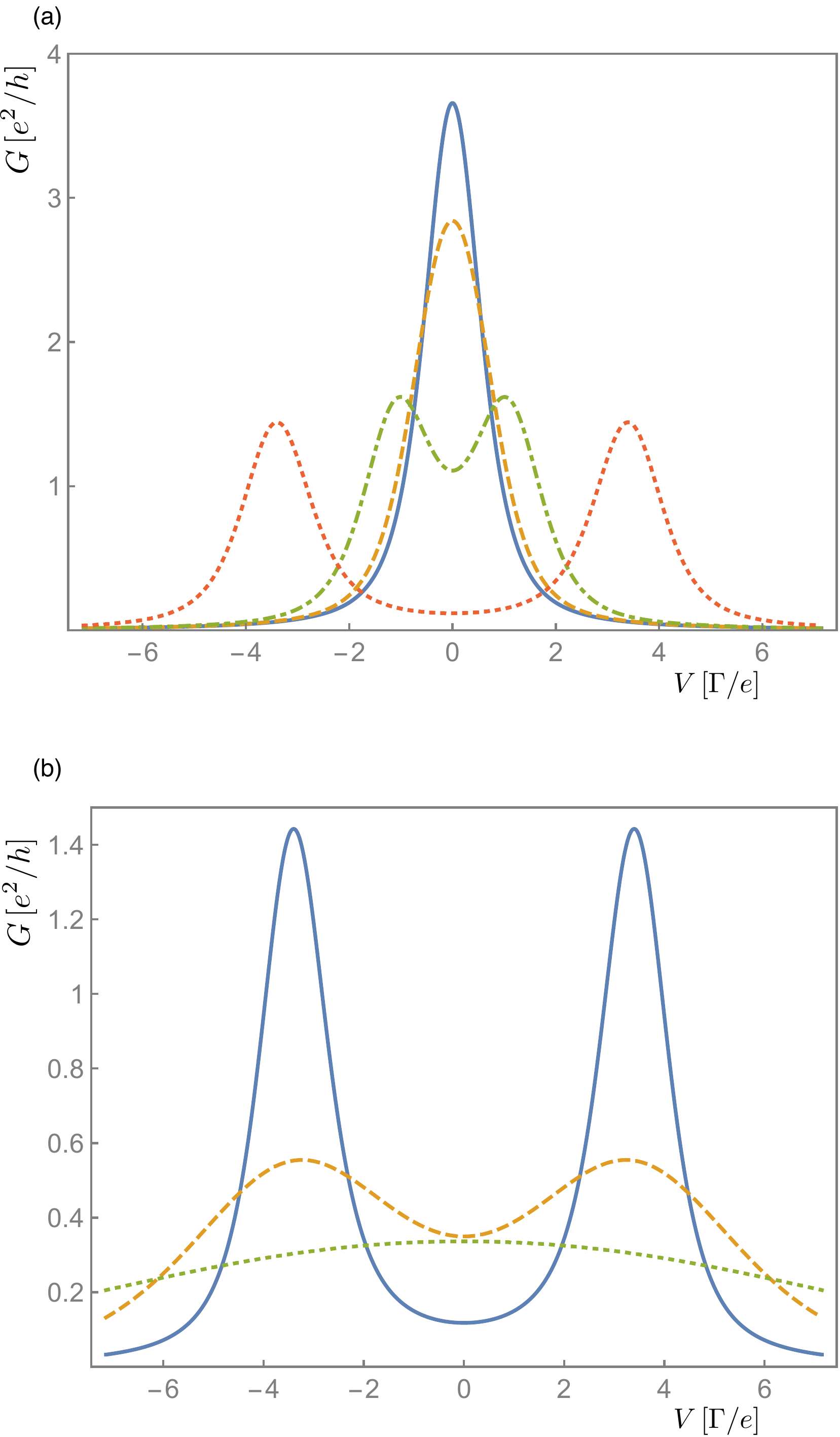}
\caption{Effect of interaction and temperature on the low bias conductance of a $4$-channel wire
  contacted to a normal lead. (a) Conductance vs.\ applied voltage
  bias at $k_BT/\Gamma=0.24$ for increasing values of the interaction
  strength: $W/\Gamma=0.03$ (full blue curve), $W/\Gamma=0.23$ (dashed
  orange curve), and $W/\Gamma=0.58$ (dot-dashed green
  curve), $W/\Gamma=1.71$ (dotted red line). Interactions induce a splitting of the  peak into a
  symmetric double peak with maxima at $eV \sim \pm 2W$ and
  renormalize the peak(s) height(s). (b) Conductance vs.\ applied
  voltage bias for $W/\Gamma=1.71$ at different temperatures:
  $k_BT/\Gamma=0.24$ (full blue curve), $k_BT/\Gamma=1.2$ (dashed
  orange curve), $k_BT/\Gamma=3.6$ (dotted green curve). Temperatures
  smears the peaks at $k_B T \sim \Gamma=3.6$, and spoils the two-peaks feature as soon as $k_BT \sim W$. In all plots the tunneling rates of the four channels are set to $\Gamma_1/\Gamma=0.43$, $\Gamma_2/\Gamma=0.31$, $\Gamma_3/\Gamma=0.20$, $\Gamma_4/\Gamma=0.06$.}
\label{fig2}
\end{figure}

\begin{figure}
\includegraphics[width=0.45\textwidth]{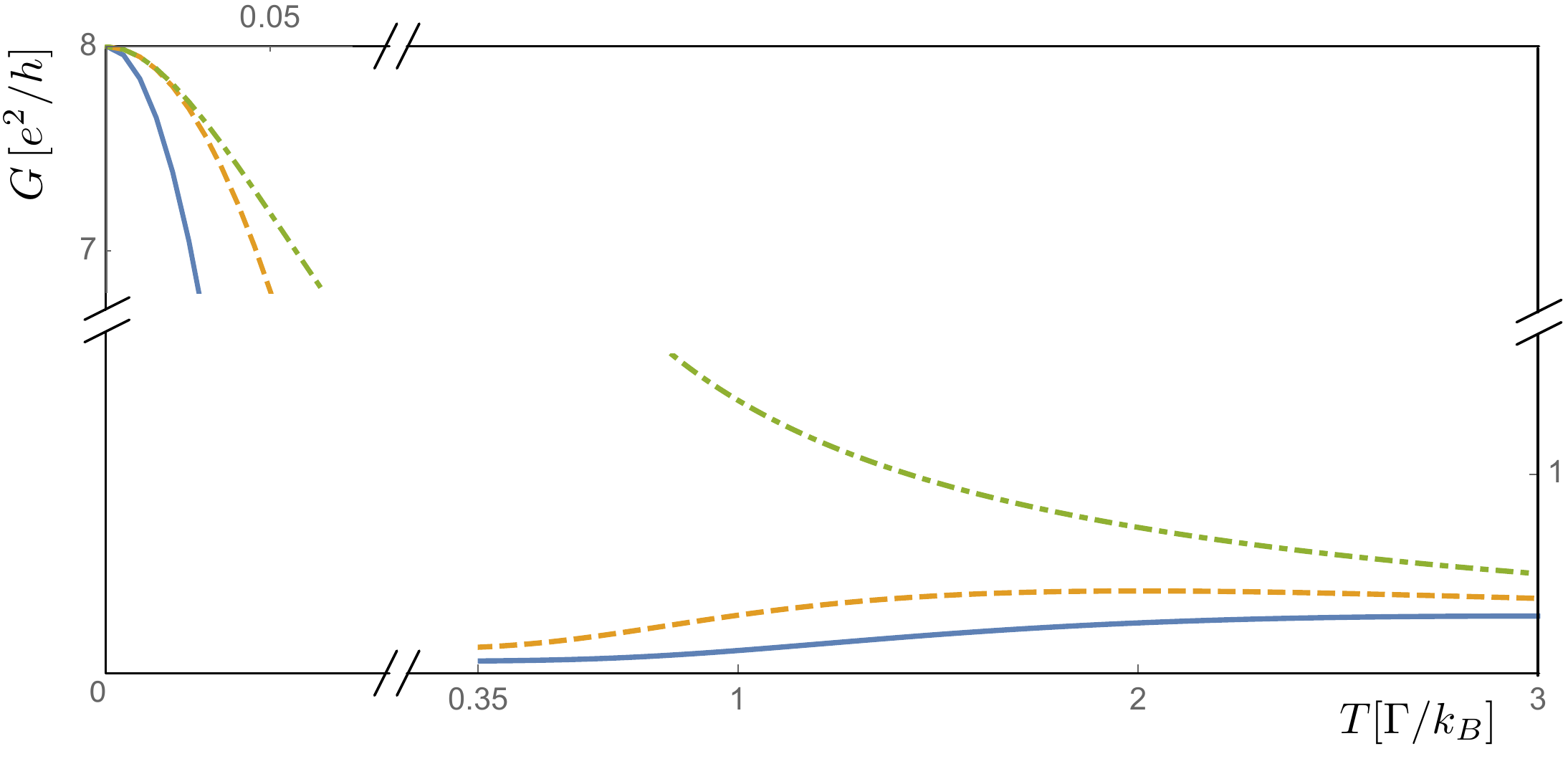}
\caption{Temperature dependence of the zero bias conductance $G(V=0)$ for different values of the interaction strength, $W/\Gamma=2.5$ (full blue curve), $W/\Gamma=1.75$ (dashed orange line), and $W/\Gamma=0$ (dot-dashed green
  curve) for high temperature. We sketch also the  low temperature behaviour ($T \ll T_K$) according to Eq.~(\ref{conductance}) (with $\alpha=1$). The reported non-monotonous behaviour of the zero-bias conductance as a function of temperature is a characteristic feature induced by interactions. The values of the tunneling rates are set to $\Gamma_1/\Gamma=0.175$, $\Gamma_2/\Gamma=0.225$, $\Gamma_3/\Gamma=0.275$, $\Gamma_4/\Gamma=0.325$.
}
\label{fig2-2}
\end{figure}

Without interactions one verifies that Eq.\ (\ref{eq:GThigh}) and (\ref{eq:GR}) reproduce the quantized conductance (\ref{eq:Gquantized}) at zero temperature, with a thermally broadened conductance peak for tempertures $T \gtrsim \min_{\alpha} \Gamma_{\alpha}$. Without interactions, the zero-bias conductance is a monotonously decreasing function of temperature.
The effects of interactions on the conductance at $T \gg T_{\rm K}$ are shown in Fig.\ \ref{fig2}. Upon increasing the interaction strength, the zero-energy peak characteristic of the non-interacting system splits into two separate peaks centered at $V \sim \pm 2W/e$. The width of the peaks is set by $\sim \max( T, \Gamma )$. Interactions concurrently reduce the height of the peak. When temperature is increased beyond the interactions strength $|W|$ the double peak structure is continuously smeared out, and a single broad peak is observed.

Interactions also affect in a characteristic way the dependence of the zero-bias conductance $G(V=0,T)$ on temperature, see Fig.~\ref{fig2-2}. 
In the weakly interacting limit,  $W \ll \Gamma$, the zero-bias conductance $G(V=0,T)$ is a monotonously decreasing function of temperature, characteristic of a single thermally-broadened conductance peak of width $\sim \Gamma$. 
Conversely, for strong interactions $W \gg \Gamma$, the conductance $G(V=0,T)$ has a maximum for $k_{\rm B} T \sim W$. Above this temperature, the conductance at finite voltage shows a single thermally broadened peak, whose height is $G(V=0,T)\propto \Gamma/k_{\rm B} T$. At lower temperatures $T<W $ a split peaks develops, leading to a zero-bias conductance that decreases when $T$ decreases, see Fig.  \ref{fig2}b, and the discussion in the preceding paragraph. As will be discussed in the next section, this behavior changes at even lower temperatures $T \lesssim T_{\rm K}$, when a Kondo resonance develops, and the zero bias conductance rises to the zero-temperature limit $G= 8 e^2/h$. 

\section{Low-temperature limit} \label{sec:5}

For strong interactions $|W| \gg \Gamma$ the perturbative result of the previous subsection predicts a zero-bias zero-temperature conductance far below the quantized limit (\ref{eq:Gquantized}). This is, however, not the true zero-temperature limit, because Kondo physics sets in at the scale $T \sim T_{\rm K}$ with
\begin{equation}
  T_{\rm K} \sim \sqrt{\Gamma W} e^{-W/\Gamma}
  \label{eq:TKondo}
\end{equation}
the corresponding Kondo temperature. To see the origin of the Kondo physics, it is instructive to follow the original problem of Sec.\ \ref{sec:model} for $|\nmajorana| = 4$ through a sequence of mappings.\cite{Meidan2014} 

The first mapping is onto a ``compactified'' Anderson impurity model.\cite{Coleman1995,Coleman1995a,Fabrizio1996} Hereto one introduces fermionic operators for pairs of Majorana states,
\be
  f_\uparrow= \frac{1}{2} (\gamma_1 +i\gamma_2),\ \ f_\downarrow= \frac{1}{2}
(\gamma_3+i \gamma_4),
\label{eq:fermioni}
\ee
and, similarly, pairwise combines the lead fermions into symmetric and antisymmetric combinations,
\begin{eqnarray}
  g_{{\rm L},\xi\uparrow} &=& \frac{1}{\sqrt{2}}( c_{\xi,1} + i c_{\xi,2} ), \\
  g_{{\rm R},\xi,\uparrow} &=& \frac{1}{\sqrt{2}}( -c_{-\xi,1}^\dag - i c_{-\xi,2}^\dag ), \\
  g_{{\rm L},\xi\downarrow} &=& \frac{1}{\sqrt{2}}( c_{\xi,3} + i c_{\xi,4} ), \\ 
  g_{{\rm R},\xi,\downarrow} &=& \frac{1}{\sqrt{2}}( -c_{-\xi,3}^\dag - i c_{-\xi,4}^\dag ).
\end{eqnarray}
In terms of these operators, the three terms $H_{\rm lead}$, $H_{\rm T}$, and $ H_{\rm int}$ contribution to the Hamiltonian become
\begin{eqnarray}
  H_{\rm lead} &=& \int \frac{d\xi}{2\pi} \sum_{{\beta={\rm L},{\rm R}} \atop \sigma = \uparrow, \downarrow} \xi g_{\beta,\xi\sigma}^\dag g_{\beta,\xi\sigma}, \nonumber \\
  H_{\rm T} &=& \int d\xi \sum_{{\beta={\rm L},{\rm R}} \atop \sigma = \uparrow, \downarrow}
  \frac{1}{\sqrt{2}} \left( V_{\sigma}  g_{\beta,\xi\sigma}^\dag f_\sigma +
  \tilde{V}_{\sigma} g_{\beta,\xi\sigma}^\dag f_\sigma^\dag +
{\rm h.c.}\right), 
  \nonumber  \\
  H_{\rm int} &=& W(2f_\uparrow^\dag f_\uparrow-1)(2f_\downarrow^\dag f_\downarrow-1),
  \label{eq:AndersonLR}
\end{eqnarray}
with the tunneling amplitudes
\begin{eqnarray}
  V_\uparrow &=& \frac{1}{2} (t_{1}+t_{2}), \ \
  V_\downarrow=  \frac{1}{2} (t_{3}+t_{4}), \nonumber \\
  \tilde{V}_\uparrow &=& \frac{1}{2} (t_{1}-t_{2}), \ \
  \tilde{V}_\downarrow=  \frac{1}{2} (t_{3}-t_{4}).
\end{eqnarray}

This mapping is followed by a transformation to symmetric and antisymmetric combinations of the lead fermions,
\begin{eqnarray}
  g_{{\rm L},\xi\sigma} &=& \frac{1}{\sqrt{2}}(g_{{\rm S},\xi\sigma} +
  g_{{\rm A},\xi\sigma}),\nonumber \\
  g_{{\rm R},\xi\sigma} &=& \frac{1}{\sqrt{2}}(g_{{\rm S},\xi\sigma} -
  g_{{\rm A},\xi\sigma}),
\end{eqnarray}
which decouples the antisymmetric modes from the Majorana end states,
\begin{eqnarray}
  H_{\rm lead} &=& \int \frac{d\xi}{2\pi} \sum_{{\beta={\rm S},{\rm A}} \atop \sigma = \uparrow, \downarrow} \xi g_{\beta,\xi\sigma}^\dag g_{\beta,\xi\sigma},\\
  H_{\rm T} &=& \int d\xi \sum_{\sigma = \uparrow, \downarrow}
  \left( V_{\sigma}  g_{{\rm S},\xi\sigma}^\dag f_\sigma +
  \tilde{V}_{\sigma} g_{{\rm S},\xi\sigma}^\dag f_\sigma^\dag +
{\rm h.c.}\right),    \nonumber 
\end{eqnarray}
In this formulation, the effective time-reversal symmetry maps into a particle-hole symmetry for the superconductor end states,
\be
  {\cal T} f_{\sigma} {\cal T}^{-1} = f_{\sigma}^{\dagger}.
\ee
For the lead fermions the action of the effective time-reversal symmetry is,
\be
  {\cal T} g_{{\rm S},\xi\sigma} {\cal T}^{-1} =
  - g_{{\rm S},-\xi,\sigma}^{\dagger},
  \label{eq:Tg}
\ee
and one easily verifies that each of the three contributions $H_{\rm lead}$, $H_{\rm T}$, and $ H_{\rm int}$ is invariant under the action of ${\cal T}$.

Without the term proportional to $\tilde V_{\sigma}$ this Hamiltonian is the standard Anderson impurity model, be it without spin rotation symmetry. The term proportional to $\tilde V_{\sigma}$ describes anomalous processes, with Andreev reflection instead of normal scattering. (Note, however, that the transformations leading to this effective Anderson model mix creation and annihilation operators, so that normal and Andreev processes in the language of the lead fermions $g_{\beta,\xi\sigma}$ not necessarily corresponds to normal or Andreev processes in the language of the original lead fermions $c_{\xi\alpha}$.) 

To find the transport properties of the model for temperatures $T\ll T_{\rm K}$ (which is where the perturbation theory of the previous Section fails), one more transformation is needed, similar to the Schrieffer-Wolff transformation of the standard Anderson impurity model,\cite{Hewson} which maps the problem to a version of the Kondo Hamiltonian,
\begin{align}
\label{compact-Kondo}
  H_{\textrm{K}}=& \int d\xi \sum_{\sigma=\uparrow,\downarrow} \xi g_{{\rm S},\xi\sigma}^\dagger g_{{\rm S},\xi\sigma} \nonumber \\ & + \sum_{\alpha=x,y,z}
[J_\alpha {\sigma}_{\alpha}(0) +\Delta_\alpha {\tau}_{\alpha}(0)]S_\alpha \, ,
\end{align}
where 
\begin{align}
& & \sigma_\alpha (0) \equiv \int d\xi d\xi' (g_{{\rm S},\xi\uparrow}^\dag,
g_{{\rm S},\xi\downarrow}^\dag) \sigma_\alpha \left( \begin{array}[c]
 c g_{{\rm S},\xi'\uparrow} \label{eq:spin}\\
g_{{\rm S},\xi'\downarrow} \end{array}\right), \\
& & \tau_\alpha (0) \equiv \int d\xi d\xi' (g_{{\rm S},\xi\uparrow}^\dag,
g_{{\rm S},\xi\downarrow}) \sigma_\alpha \left( \begin{array}[c]
 c g_{{\rm S},\xi'\uparrow} \label{eq:p-h}\\
g_{{\rm S},\xi'\downarrow}^\dag \end{array}\right),
\end{align}
$\sigma_\alpha$, $\alpha=x,y,z$ being the Pauli matrices, and
\begin{align}
  J_x =& ( t_2 t_4 +t_1 t_3)/W, \nonumber \\
  J_y =& (t_2 t_3 + t_1 t_4)/W, \nonumber \\
  J_z =& (t_1 t_2+t_3 t_4)/W, \\
  \Delta_x =& ( t_2 t_4 -t_1 t_3)/W, \nonumber \\
  \Delta_y =& (t_2 t_3 - t_1 t_4 )/W, \nonumber \\
  \Delta_z =& (t_1 t_2 -t_3 t_4)/W.
\end{align}

The Hamiltonian (\ref{compact-Kondo}) is known as the ``compactified two-channel Kondo model''\cite{Coleman1995,Coleman1995a,Fabrizio1996} and it has a phenomenology that closely resembles the standard two-channel Kondo model. In particular, it has a non-Fermi liquid fixed point if the ``normal'' and ``anomalous'' couplings $J_{\alpha}$ and $\Delta_{\alpha}$ are equal, whereas it flows to the Fermi-liquid fixed point of the conventional single-channel Kondo model otherwise. In our case, the requirement that all four tunneling parameters $t_{\alpha}$ be positive ensures that the ``normal'' couplings $J_x$, $J_y$, and $J_z$ in the Hamiltonian (\ref{compact-Kondo}) dominate. Therefore, at low temperature and voltage the model (\ref{compact-Kondo}) flows to the isotropic Kondo singlet fixed point, where it is described by a unitary scattering matrix $s=-\openone_2$ for the symmetric mode of the lead fermions,\cite{Hewson} without Andreev processes. Retracing the transformations of the present Section one then finds that at zero temperature and voltage the original multichannel Majorana model of Sec.\ \ref{sec:model} is characterized by a unitary reflection matrix which takes the simple form\cite{Meidan2014} $r_{\rm ee} = r_{\rm hh} = 0$, $r_{\rm he} = r_{\rm eh} = \openone_{4\times 4}$. The conductance at zero temperature in linear response readily follows as $G = (2e^2/h) \mbox{tr}\, r_{\rm eh}r_{\rm eh}^{\dagger} = 8 e^2/h $, whereas the shot noise vanishes, $F=0$. 

At finite temperature, inelastic scattering renders the scattering matrix nonunitary. To study the corrections to the conductance of the interacting Majorana wire at finite temperature and voltage, we first discuss the transport of the compactified Anderson model, then discuss the implications of the mapping to our model of interest. 

\subsection{Transport properties of the compactified Anderson model}

To analyze the perturbative corrections around this point it is convenient to rewrite the model (\ref{compact-Kondo}) in real space on a discrete lattice as
\begin{equation}\label{s-d_model}
H=\sum_\sigma \sum_{i,j \ge 0} t_{i,j} g^\dagger_{i,\sigma} g_{j,\sigma} +
[J \vec{\sigma}(0) +\Delta \vec{\tau}(0) ]\cdot \vec{S},
\end{equation}
where the hopping amplitude $t_{i,j}$ is nonzero for nearest neighbors only, and with the local electron spin and particle-hole degree of freedom at site $i=0$ given by $\vec{\sigma}(0) =\{\sigma_x (0),\sigma_y(0),\sigma_z(0) \}$, with the definitions in Eqs.~(\ref{eq:spin},\ref{eq:p-h}) and  $\vec{S}=\{ S_x,S_y,S_z \} $. We have suppressed the subscript ``S'' referring to the symmetric mode.

Since at zero temperature the antiferromagnetic coupling $J > 0$ is the dominant term, corrections around the singlet fixed point can be obtained as a perturbation in the tunneling $t_{i,0}$, {\em i.e.}, at the strong coupling limit $D/J \ll1$, where $D \sim t_{i,i+1}$ is the band width for the lead fermions. This is done by tracing out the local degrees of freedom of the bound state of the spin and electron on site $0$, in order to obtain an effective theory for a free fermionic lead.\cite{Nozieres1974} Without the anomalous term proportional to $\Delta$ the leading corrections in a small-$D/J$ expansion are a correction to the hopping amplitude $t_{0,1}$ of order $D^3/J^2$ and a repulsive interaction $\propto g_{1,\uparrow}^\dagger g_{1,\downarrow}^\dagger g_{1,\downarrow} g_{1,\uparrow}$ of order $D^4/J^3$. The main difference between the present analysis and that of the standard single channel Kondo problem is the appearance of the ``anomalous'' term proportional to $|\Delta| \ll J$ in the perturbative expansion of the effective Hamiltonian.

Accounting for the presence of a nonzero anomalous term perturbatively, we find to leading order $D^2 \Delta/J^2$ no corrections quadratic in the operators $g_{i,\sigma}$ and $g^{\dagger}_{i,\sigma}$. The absence of quadratic corrections can also follows from the observation that all six possible local terms $2 g_{1,\uparrow}^{\dagger} g_{1,\uparrow} - 1$, $2 g_{1,\downarrow}^{\dagger} g_{1,\downarrow} - 1$, $g_{1,\uparrow}^{\dagger} g_{1,\downarrow} + g_{1,\downarrow}^{\dagger} g_{1,\uparrow}$, $i g_{1,\uparrow}^{\dagger} g_{1,\downarrow} - i g_{1,\downarrow}^{\dagger} g_{1,\uparrow}$, $g_{1,\uparrow}^{\dagger} g_{1,\downarrow}^{\dagger} + g_{1,\downarrow} g_{1,\uparrow}$, and $i g_{1,\uparrow}^{\dagger} g_{1,\downarrow}^{\dagger} -i g_{1,\downarrow} g_{1,\uparrow}$ are antisymmetric under application of the effective time-reversal operation ${\cal T}$, see Eq.\ (\ref{eq:Tg}), whereas corrections to hopping amplitudes require higher orders in $D/J$. Thus, the leading corrections from the anomalous term involve corrections to the prefactors of  the hopping amplitude $t_{0,1}$ and the repulsive on-site interaction $\propto g_{1,\uparrow}^\dagger g_{1,\downarrow}^\dagger g_{1,\downarrow} g_{1,\uparrow}$. These corrections to the prefactors are a factor $\sim \Delta/J$ smaller than the prefactors in the absence of anomalous terms, at $\Delta=0$. Since these correction terms affect the {\em deviations} from the zero-temperature limit, we conclude that the inclusion of the anomalous term leads to small (order $\Delta/J$) modifications of the finite-temperature and finite-voltage corrections to the zero-temperature limit, whereas it does not affect the zero-temperature limit itself. In particular, inclusion of the anomalous term leads to a correction of (relative) order $\Delta/J$ to the Kondo temperature $T_{\rm K}$.

\subsection{Transport properties of interacting chain }
\label{signatures}
\paragraph{Current} 

We now return to the original problem of the multichannel Majorana chain of Sec.\ \ref{sec:model}. We start by considering the current operator (\ref{eq:current}) in the multi-channel Majorana chain, as well as current operators in the left and right leads (defined as the rate of charge flow {\em into} the dot) in the Anderson model, Eq.\ \eqref{eq:AndersonLR},
\begin{align}
\tilde{I}_{\rm L} &=\frac{ie}{\hbar\sqrt{2}}
  \int d\xi \sum_{\sigma}g_{{\rm L},\xi\sigma}^{\dagger}
  \left(V_\sigma  f_{\sigma} +\tilde{V}_\sigma f_\sigma^\dagger \right)+\textrm{h.c.,} \nonumber \\
\tilde{I}_{\rm R} &=\frac{ie}{\hbar\sqrt{2}} \int d\xi \sum_{\sigma}g_{{\rm R},\xi\sigma}^{\dagger}\left(V_\sigma  f_{\sigma} +\tilde{V}_\sigma f_\sigma^\dagger \right)+\textrm{h.c.}.
\end{align}
Using the mapping between the four-channel Majorana chain and the Anderson model, we can express  $\tilde{I}_{\rm R}$ and $\tilde{I}_{\rm L}$ in terms of the current $I$ in the multichannel Majorana model as 
\be
  I = \tilde{I}_{\rm L} - \tilde{I}_{\rm R}.
  \label{eq:IILR}
\ee
Furthermore the distribution funcions, $\langle c_{\xi,\alpha}^\dagger c_{\xi,\alpha}\rangle=f(\xi-eV)$, of the voltage biased lead in the original Majorana model are mapped onto the distribution function of the left and right leads as
\begin{align}
\langle g_{L,\xi\sigma}^\dagger g_{L,\xi\sigma} \rangle =& f(\xi-eV), \nonumber \\  \langle g_{R,\xi\sigma}^\dagger g_{R,\xi\sigma} \rangle =& f(\xi+eV),
\label{eq:distribuzioni}
\end{align}
with all other correlations vanishing. Note that Eq.\ (\ref{eq:distribuzioni}) implies that in the Anderson model the voltage bias between the left and right leads is $2V$.
Equations (\ref{eq:IILR}) and (\ref{eq:distribuzioni}) show explicitly how the current $I$ in the multichannel Majorana model can be calculated from the currents $\tilde I_{\rm R}$ and $\tilde I_{\rm L}$ in the Anderson model (\ref{eq:AndersonLR}).

At zero temperature and bias, the Anderson model is described by the Kondo fixed point, without anomalous terms, so that one has $\langle \tilde{I}_{\rm L} \rangle = - \langle \tilde{I}_{\rm R} \rangle = 2 e^2 (2V)/h$,\cite{Glazman1988,Ng1988} from which it follows that 
\be
  \langle I \rangle = 8 e^2 V/h
  \label{eq:I00}
\ee
at zero temperature and zero bias.
At finite temperature and/or voltage, but without the anomalous terms, the expectation values $\langle \tilde{I}_{\rm L} \rangle$ and $\langle \tilde{I}_{\rm R} \rangle$ are suppressed by a factor $1 - (\alpha/T_{\rm K})^2 [(2 e V)^2/2 + \pi T^2]$,\cite{Sela2006,Mora2009} where $\alpha$ is a coefficient of order unity and $T_{\rm K}$ the Kondo temperature, see Eq.\ (\ref{eq:TKondo}). For the four-channel Majorana model this implies that the conductance is 
\be
  G=\frac{8e^{2}}{h}\left[1-\left(\frac{\alpha}{T_{K}}\right)^{2}\left(\frac{1}{2}(2eV)^{2}+\pi T^{2}\right)\right].
  \label{conductance}
\ee
Following the discussion of the previous subsection, inclusion of the anomalous term only quantitatively affects the finite-temperature/voltage corrections, these corrections being parametrically small in the ratio $\Delta/J$. This means that the result (\ref{conductance}) continues to hold in the presence of the anomalous terms, though with a small change to the numerical constant $\alpha$ or, alternatively and equivalently, to the definition of $T_{\rm K}$.

\paragraph{Shot noise}

The zero-frequency shot noise power $S$ is
\be
  S = \lim_{\tau \to \infty} \frac{1}{\tau} \langle \delta Q(\tau)^2 \rangle,
\ee
where
\be
  \delta Q(\tau) = \int_0^{\tau} d\tau' [I(\tau') - \langle I \rangle]
\ee
is the fluctuation of the transported charge for a time interval of duration $\tau$. Using the relation (\ref{eq:IILR}) one has $\delta Q(\tau) = \delta \tilde{Q}_{\rm L}(\tau) - \delta \tilde{Q}_{\rm R}(\tau)$, with
\be
  \delta \tilde{Q}_{\rm L,R}(\tau) =
  \int_0^{\tau} d\tau' [\tilde{I}_{\rm L,R}(\tau') - \langle \tilde{I}_{\rm L,R} \rangle].
\ee
For the zero-frequency shot noise this implies
\be
  S = \lim_{\tau \to \infty} \frac{1}{\tau} \left\langle 2 \delta \tilde Q_{\rm R}(\tau)^2 + 2 \delta \tilde{Q}_{\rm L}(\tau)^2 - \delta \tilde{Q}_{\rm imp}(\tau)^2 \right\rangle,
  \label{eq:Stilde}
\ee
with $\delta \tilde{Q}_{\rm imp}(\tau) = \delta \tilde{Q}_{\rm L}(\tau) + \delta \tilde{Q}_{\rm R}(\tau)$
the fluctuation of the charge on the impurity site in the Anderson model of Eq.\ (\ref{eq:AndersonLR}).

Without the anomalous terms in Eq.\ (\ref{eq:AndersonLR}), the charge on the impurity site is bounded, so that the third term in Eq.\ (\ref{eq:Stilde}) vanishes after taking the limit $\tau \to \infty$. The first two terms in Eq.\ (\ref{eq:Stilde}) are equal and correspond to the zero-frequency shot noise $\tilde S$ in the Anderson model, so that we conclude
\begin{align}
\label{shot noise}
  S = 4 \tilde{S} = \frac{8e^{2}}{h}e|V|\,\frac{10}{3}\alpha^2\left(\frac{2eV}{T_{K}}\right)^{2},  
\end{align}
where, in the second line, we substituted the known result for the zero-frequency shot noise power $\tilde S$ in  the Anderson model, see Refs.\ \onlinecite{Sela2006,Mora2009}. 

Following common practice in the study of the Kondo problem, we define a Fano factor $F'=S/2e |\delta I|$ as the ratio of the shot noise power $S$ and the {\em difference} $\delta I = I - 8 e^2 V/h$ of the zero-temperature current $I$ at finite bias $V$ and the zero-temperature current at vanishing bias (\ref{eq:I00}). Such a definition is necessary, because the zero-temperature zero-bias current, used in the standard definition of the Fano factor, is noiseless. Combining the results obtained above, we find
\be
  F' = \frac{10}{3}.
\ee
The Fano factor $F$ describes the noise properties of particles that are normally reflected off the superconducting interface. It may be interpreted as the effective charge (in units of $e$) of particles normally reflected in the four-channel Majorana wire, seen against the background of the current carried by perfectly Andreev-reflected electrons.

As in the discussion of the average current, to leading order at low temperature and/or bias the inclusion of the anomalous terms in Eq.\ (\ref{eq:AndersonLR}) only quantitatively modifies the effective model parameters, but not the form of the results. Hence, its effect is a modification of the coefficient $\alpha$, which is small if $\Delta/J$ is small. However, the result for the Fano factor $F'$, which does not depend on the precise value of $\alpha$, does not change.

The expressions \eqref{conductance} for the current $I$ and \eqref{shot noise} for the zero-frequency shot noise power $S$ should be compared to the current and shot noise of a non-interacting four-channel Majorana wire (see App.\ \ref{appendix3}),
\begin{align}
I =& \frac{8 e^2}{h}V\left[1-\frac{1}{3} \left(\frac{eV}{\Gamma}\right)^2\right],
  \label{eq:Inonint}
\\
S =&  \frac{8e^2}{h}e|V| \frac{4}{3} \left(\frac{eV}{\Gamma}\right)^2,
  \label{eq:Snonint}
\end{align}
which has Fano factor
\be
  F' = 2,
\ee
corresponding to an effective charge $e^* = 2 e$ of normally reflected electrons (against the background of Andreev reflected electrons). We therefore conclude that the fano factor $F'$ constitutes a clear experimental signature of the emergent topological Kondo resonance. 


\section{Time reversal symmetry breaking}
\label{sec:6}

The low-temperature transport properties derived in the previous section are unstable against perturbations that break the effective time-reversal symmetry. In the absence of time-reversal symmetry, quadratic terms coupling the Majorana end states are possible. We here consider the effect of a generic time-reversal-symmetry breaking perturbation.

If the number of Majorana end states $-3 \le \nmajorana \le 3$, the low-energy theory is unaffected by the presence of interactions, and the analysis of time-reversal breaking perturbations essentially coincides with that of the non-interacting system.\cite{GKellsPRB2012a} The effect of generic time-reversal-symmetry breaking terms is to pairwise gap out Majorana end states, leaving behind a single zero-energy Majorna end-state if $\nmajorana$ is odd, and none otherwise. Correspondingly the zero-temperature conductance will show a quantized value $2 e^2/h$ at zero bias in the case of odd $\nmajorana$, whereas the zero-bias conductance is zero if $\nmajorana$ is even. The gapped-out Majorana end states give rise to (quantized) conductance peaks at finite voltage; the effect of a finite temperature is to thermally smear the conductance.

The case $\nmajorana = \pm4$ demands a more careful analysis. For concreteness we consider $\nmajorana=4$, with a generic time-reversal-breaking perturbation
\begin{align}
\label{eq:TRB-pert}
H_{\rm TRB} = i \sum_{\alpha,\beta=1}^{4} Y_{\alpha,\beta} \gamma_\alpha \gamma_\beta ,
\end{align}
with real coefficients $Y_{\alpha,\beta} = -Y_{\beta,\alpha}$. In the specific implementation of a spinless $p$-wave superconducting quantum wires with multiple transverse modes in the topological phase, such time-reversal breking terms emerge from transverse superconducting pairing.\cite{GKellsPRB2012a} Following the mapping to the compactified Anderson model in Section~\ref{sec:5},
after the introduction of the fermionic degrees of freedom $f_\uparrow$, $f_\downarrow$ in Eq.~(\ref{eq:fermioni}), the perturbation in Eq.~(\ref{eq:TRB-pert})  takes the form 
\begin{align}
 H_{\rm TRB} = \mbox{} &   (B_z+ \varepsilon_0) f_\uparrow^\dagger f_\uparrow 
  - (B_z -\varepsilon_0) f_\downarrow^\dagger f_\downarrow 
  \nonumber \\ & 
  + (B_x-iB_y) f_\uparrow^\dagger f_\downarrow + (B_x+iB_y)
  f_\downarrow^\dagger f_\uparrow 
  \nonumber \\ & 
  + (D_x-iD_y) f_\uparrow^\dagger f_\downarrow^\dagger 
  + (D_x+iD_y) f_\downarrow f_\uparrow.
\end{align}
where the six parameters $B_x$, $B_y$, $B_z$, $\varepsilon_0$, $D_x$, and $D_y$ are expressed in terms of the original six parameters $Y_{\alpha,\beta}$. In the language of the Anderson model, the parameters $B_x$, $B_y$, and $B_z$ may be interpreted as an external magnetic field, $\varepsilon_0$ is interpreted as a potential that drives the model away from the symmetric point, and $D_x$ and $D_y$ correspond to a local superconducting pairing.

As in the previous Section, without loss of generality we assume that the interaction is repulsive $W>0$. Then the two ground states of the isolated dot in the absence of the time-reversal-breaking perturbation are the two states $| \uparrow \rangle $, $| \downarrow \rangle$ of a singly occupied impurity site. The two excited impurity states correspond to the empty and doubly-occupied impurity site, $\vert 0 \rangle$, $\vert \uparrow, \downarrow \rangle$. The effect of the terms proportional to $B_x$, $B_y$, and $B_z$ is to split the degeneracy of the ground state, while the superconducting and overall energy terms ($D_i,\epsilon_0$) spit the degeneracy of the excited state. The latter perturbations (to leading order) have no effect on the ground state; they drive the compactified Anderson model away from the symmetric point, and become irrelevant when the model is mapped to the corresponding compactified Kondo problem.\cite{Meidan2014} The only relevant contribution of the time-reversal-breaking perturbation is therefore from to the terms proportional to $B_x$, $B_y$, and $B_z$. Since the low-energy fixed point of model without time reversal symmetry breaking is the isotropic compactified Kondo model, we can generally assume $B_x=B_y=0$, $B_z=B$. Thus, the model in the presence of a time-reversal-symmetry breaking perturbation can be ultimately mapped, at low energy, onto the Anderson model with a Zeeman-split ground state. We are interested in the leading order correction to the conductance due to this splitting.

The effect of a magnetic field on the standard Anderson model is well known.\cite{Nozieres1974} At zero temperature, a weak magnetic field suppresses the conductance by a factor $[1 - (B/T_{\rm K})^2]$. Hence, a time-reversal-symmetry-breaking perturbation leads to a similar suppression of the zero-temperature conductance of the four-channel Majorana chain,
\be
  G = \frac{8 e^2}{h} \left[ 1 - \left( \frac{B}{T_{\rm K}} \right)^2 \right].
\ee
The corresponding correction for the non-interacting system in the presence of a time reversal symmetry breaking term $B_z = B $ is 
\be
  G = \frac{8 e^2}{h} \left[ 1 - \left( \frac{B}{2\Gamma} \right)^2 \right].
\ee
where $\Gamma = 2\pi|t_\alpha|^2 = 2\pi|t|^2$ is the level broadening for a channel independent tunneling amplitude.~\cite{Note4}
Since generically $T_{\rm K} \ll \Gamma$, cf.\ Eq.\ (\ref{eq:TKondo}), in the interacting Majorana wire the corrections from a time-reversal-symmetry-breaking perturbation are larger than without interactions.

\section{Conclusions}
\label{sec:conclusions}

In this manuscript we have systematically analyzed the effect of 
residual electron-electron interactions on the transport
properties of a topological superconducting wire with an effective
time reversal symmetry, contacted to an external lead. Such a wire
is known to have a $\mathbb{Z}_8$ topological classification, where
the topological index $\nmajorana$ counts the number of Majorana end 
states, negative $\nmajorana$ corresponding to end states that are odd 
under time reversal. Nontrivial interaction effects exist for the 
case $|\nmajorana|=4$ ($\mbox{mod}\, 8)$ only, since other values of $\nmajorana$ 
are effectively described by a non-interacting theory. 

For $|\nmajorana|=4$ interactions induce a Kondo-like correlated state.
At tempertures $T$ much larger than the associated Kondo temperature $T_{\rm K}$, 
interactions lead to a splitting of the zero-bias conductance peak
characteristic of the non-interacting system and, hence, a slight
decrease of the zero-bias conductance for temperatures
$\Gamma \ll T \ll W$ ($\Gamma$: tunnel coupling to the normal-metal lead, $W$: interaction energy). 
The formation of a Kondo resonance at temperatures 
$T \ll T_{\rm K}$, however, leads to an increase of the conductance 
with decreasing temperature and restores a 
zero-temperature and zero-bias conductance $G = 8 e^2/h$
that is indistinguishable
from the non-interacting case.\cite{Meidan2014}
For comparison: Without interactions the zero-bias conductance shows a monotonous increase upon decreasing the temperature. Although the zero-temperature
zero-bias conductance does not distinguish between the non-interacting and interacting scenarios, in the Kondo regime $T \ll T_{\rm K}$ differences between the two cases exist at low but finite temperature or bias differences. 
The most striking such difference is
the Fano factor $F'$, which describes the ratio of the shot noise
power and the normal-reflected current. We found $F' = 10/3$ in the interacting case, whereas $F' = 2$ without interactions. This difference marks another
clear experimentally detectable difference between the scenarios without
and with interactions.

The anomalous Fano factor $F'$ for the case $|n| = 4$ was obtained via a mapping to the standard Anderson impurity model. That mapping involves a factor two for the Fano factor, and the Fano factor $F' = 10/3$ follows from the anomalous Fano factor $5/3$ reported for the Anderson impurity model.\cite{Sela2006,Mora2009} The derivation of the original result requires the presence of particle-hole symmetry, which is a good approximate symmetry in the setting of the Anderson impurity model. Particle-hole symmetry is manifest in the interacting Majorana wire we consider here, so that the present system may be an interesting alternative to observe the anomalous Fano factor associated with the Kondo effect.

In Ref.\ \onlinecite{Meidan2014}, we argued that there are two possible
Fermi liquid fixed points describing the interacting Majorana wire with 
$|\nmajorana| = 4$. One of these is reminiscent of the non-interacting
system with $\nmajorana = 4$, the other one derives from the non-interacting
system with $\nmajorana = -4$. For the transport properties we consider
here, conductance and shot noise, one finds the same low-temperature
and low-bias results in these two cases. Our analysis is not
valid for the boundary between the ``$\nmajorana = 4$'' and 
``$\nmajorana = -4$'' parameter regions, where the low-energy properties
of the system are described by a (non-Fermi liquid) two-channel Kondo 
fixed point. However, the two-channel Kondo scenario requires fine
tuning of system parameters; The mapping to the single-channel
Kondo model used here is valid for generic parameter choices.

Practical realizations of Majorana wires have been proposed for hybrid systems involving topological-insulator edges,\cite{Fu2008} semiconductor wires,\cite{Lutchyn2010,Oreg2010} ferromagnetic structures,\cite{Duckheim2011,Chung2011}, or arrangements of ferromagnetic atoms.\cite{Choy2011,Kjaergaard2012,Martin2012,Nadj2013} Of these, the topological insulator platform is strictly one-dimensional by nature, which makes it unsuitable for multichannel wire realizations. A realization with semiconductor wires requires a Zeeman energy exceeding the splitting between transverse subbands,\cite{Rieder2012} which can be achieved for wide nanowires. On the other hand, ferromagnetic nanowires have multiple channels as soon as they are more than a few atoms in cross section. In all these proposed realizations, the effective time-reversal symmetry is not an exact symmetry, although it is a good approximate symmetry if the wire width $W$ is much smaller than the coherence length $\xi$ of the proximity-induced superconducting phase.\cite{Tewari2011,GKellsPRB2012a} The interactions required to observe the Kondo-like physics discussed here (interaction strength larger than the energy scale associated with the breaking of the effective time-reversal symmetry) arise from charge fluctuations inherent to Andreev bound states, see, {\em e.g.}, Ref.\ \onlinecite{Meidan2014}. Since the presence of the superconductor effectively screens interactions in a normal-metal wire in its proximity, the interaction strength may be enhanced if the end of the normal-metal wire (which is where the Majoranas reside) is not covered by a superconducting material.

\acknowledgments

We thank Leonid Glazman, Felix von Oppen, Falko Pientka, and Maresa Rieder for discussions. P. W. B. acknowledges support  by the Alexander von Humboldt Foundation in the framework of the Alexander von Humboldt Professorship, endowed by the Federal Ministry of Education and Research, and D. M. acknowledges  
support from the  Israel Science Foundation (Grant No. $ 737/14$) and 
from the European Union's Seventh Framework Programme (FP7/2007-2013) under Grant No. 631064.
\begin{widetext}

\appendix

\section{Majorana's Green's functions at high temperature.}
\label{appendix2}
 We present here the explicit calculation of the Majorana end-states
 Green's functions in the high temperature limit.
We are ultimately interested in the retarded Green's function in the energy
domain. We start form the corresponding Green's functions in time and
define  
\begin{eqnarray*}
F_{\alpha;\beta}(t) & = & -i\theta(t)\langle\left\{
  \gamma_{\alpha}(t),\gamma_{\beta}\right\} \rangle ,\\
J_{\mu<\nu<\zeta;\beta}(t) & = & -i\theta(t)\langle\left\{ \gamma_{\mu}(t)\gamma_{\nu}(t)\gamma_{\zeta}(t),\gamma_{\beta}\right\} \rangle.
\end{eqnarray*}

Repeatedly applying the equations of motions for the above functions,
and neglecting Majorana-Leads 4-operators correlations,
we obtain a closed set of equations.\cite{Meir1991} In the frequency
domain they read
\begin{eqnarray*}
(\omega+i\eta)\, F_{\alpha;\beta}(\omega) & = & 2\delta_{\alpha,\beta}-i\, t_{T,\alpha}^{2}\tilde{\Sigma}-2\sum_{\mu<\nu<\zeta}(\tilde{W}_{\alpha,\mu,\nu,\zeta}\, J_{\mu<\nu<\zeta;\beta}(\omega)),\\
(\omega+i\eta)\, J_{\mu<\nu<\zeta;\beta}(\omega) & = & 2(\delta_{\beta,\zeta}\langle\gamma_{\mu}\gamma_{\nu}\rangle-\delta_{\beta,\nu}\langle\gamma_{\mu}\gamma_{\zeta}\rangle+\delta_{\beta,\mu}\langle\gamma_{\nu}\gamma_{\zeta}\rangle)\\
 &  & -i\,(t_{T,\mu}^{2}+t_{T,\nu}^{2}+t_{T,\zeta}^{2})\,\tilde{\Sigma}_{\,}J_{\mu<\nu<\zeta;\beta}(\omega)\\ 
&  & -2\sum_{\alpha}\tilde{W}_{\alpha,\mu,\nu,\zeta\,}F_{\alpha;\beta}(\omega),
\end{eqnarray*}
with
\begin{eqnarray*}
\tilde{\Sigma} & = & 2i\,\sum_{h}\left(\frac{1}{\omega+i\eta-\epsilon_{h}}+\frac{1}{\omega+i\eta+\epsilon_{h}}\right),\\
\tilde{W}_{\alpha,\mu,\nu,\zeta\,} & = & W_{\alpha,\mu,\nu,\zeta}-W_{\mu,\alpha,\nu,\zeta}+W_{\mu,\nu,\alpha,\zeta}-W_{\mu,\nu,\zeta,\alpha}.
\end{eqnarray*}
Here $W_{\alpha,\mu,\nu,\zeta}\neq0$ only when $\alpha<\mu<\nu<\zeta$.
With four channels the only nonvanishing entry is $W_{1,2,3,4}=W$.
This gives 
\begin{eqnarray*}
F_{\alpha;\beta}(\omega) & = & 2u_{\alpha}(\omega)\delta_{\alpha,\beta}+4M_{\alpha;\beta}(\omega),\\
u_{\alpha}(\omega) & = & \frac{1}{\omega+i\tilde{\Sigma}t_{\alpha}^{2}-\sum_{\mu<\nu<\zeta}\frac{4\tilde{W}_{\alpha,\mu,\nu,\zeta}^2}{\omega+i\tilde{\Sigma}(t_{T,\mu}^{2}+t_{T,\nu}^{2}+t_{T,\zeta}^{2})}},\\
M_{\alpha;\beta}(\omega) & = &- \sum_{\mu<\nu<\zeta}\frac{\tilde{W}_{\alpha,\mu,\nu,\zeta}}{\omega+i\tilde{\Sigma}(t_{\mu}^{2}+t_{\nu}^{2}+t_{\zeta}^{2})}\left(\delta_{\beta,\zeta}\langle\gamma_{\mu}\gamma_{\nu}\rangle-\delta_{\beta,\nu}\langle\gamma_{\mu}\gamma_{\zeta}\rangle+\delta_{\beta,\mu}\langle\gamma_{\nu}\gamma_{\zeta}\rangle\right).
\end{eqnarray*}

The averages of the two-points-same-time Majorana operators have to
be evaluated in terms of some temperature occupation function of the
end-states. Note that, since in $\tilde{W}_{\alpha,\mu,\nu,\zeta}$ $\alpha\neq\mu,\nu,\zeta$
and $\beta$ appears in delta-functions with $\mu,\nu,\zeta$, $M_{\alpha;\beta}\neq0$
only for $\alpha\neq\beta$. Here we include the interaction also in determining the
energy levels of the local Majorana system. We know that the interaction
produces a level splitting into two doubly degenerate spaces at energy
$\pm W$. Assuming thermal occupation of the Majorana's levels, we
have the following density matrix to average the Majorana's two-points-function:
\[
\rho=\frac{1}{2}\left(\begin{array}{cccc}
f(W) & 0 & 0 & 0\\
0 & f(W) & 0 & 0\\
0 & 0 & f(-W) & 0\\
0 & 0 & 0 & f(-W)
\end{array}\right),
\]
where the matrix is written in the basis $\vert1,0\rangle,\:\vert0,1\rangle,\,\vert1,1\rangle\,\vert0,0\rangle$,
$W>0$, and $f(\epsilon)\equiv(1+e^{\beta\epsilon})^{-1}$ is the
Fermi distribution function. Note that $f(\epsilon)+f(-\epsilon)=1$
guarantees $\textrm{Tr}\rho=1$. We obtain 
\[
\langle\gamma_{\mu}\gamma_{\nu}\rangle\equiv\textrm{Tr}\left[\gamma_{\mu}\gamma_{\nu}\rho\right]=0\,\textrm{for }\mu\neq\nu.
\]
Hence 
\[
M_{\alpha;\beta}(\omega)=0.
\]
The only nonzero contributions to the Majorana's green's function
are 
\begin{eqnarray*}
F_{1;1} & = & \frac{2}{\omega+it_{T,1}^{2}\tilde{\Sigma}-\frac{4W^2}{\omega+i(t_{T,2}^{2}+t_{T,3}^{2}+t_{T,4}^{2})\tilde{\Sigma}}}\\
F_{2;2} & = & \frac{2}{\omega+it_{T,2}^{2}\tilde{\Sigma}-\frac{4W^2}{\omega+i(t_{T,1}^{2}+t_{T,3}^{2}+t_{T,4}^{2})\tilde{\Sigma}}}\\
F_{3;3} & = & \frac{2}{\omega+it_{T,3}^{2}\tilde{\Sigma}-\frac{4W^2}{\omega+i(t_{T,1}^{2}+t_{T,2}^{2}+t_{T,4}^{2})\tilde{\Sigma}}}\\
F_{4;4} & = & \frac{2}{\omega+it_{T,4}^{2}\tilde{\Sigma}-\frac{4W^2}{\omega+i(t_{T,1}^{2}+t_{T,2}^{2}+t_{T,3}^{2})\tilde{\Sigma}}}.
\end{eqnarray*}

Note that, importantly, the above procedure can be repeated for the
functions $-i\theta(t)\langle\ \gamma_{\alpha}(t),\gamma_{\beta}
\rangle$,  $-i\theta(t)\langle\ \gamma_{\alpha},\gamma_{\beta}(t)
\rangle$, the only difference being the apparence of $\langle
\gamma_\alpha \gamma_\beta \rangle$ instead of the
$\delta_{\alpha,\beta}$ at same time points. Therefore the matrix
structure of all the equations in the channel indices is preserved and, since $\langle
\gamma_\alpha \gamma_\beta \rangle \propto \delta_{\alpha,\beta}$, all
the Green's functions $G^{>(<)}_{\alpha,\beta} \propto \delta_{\alpha,\beta}$.

\section{Shot noise for non-interacting Majoranas}
\label{appendix3}

In the absence of interactions the four channels are completely decoupled and we may calculate the shot noise of each channel separately. 
The expression for the shot noise in a single-channel normal-metal--superconductor junction is\cite{Jong1994,Anantram1996}
\begin{align}
  S =& \mbox{} \frac{2 e^2}{h} \int_0^{\infty} d\varepsilon
  \left\{ {\cal I}_{\rm ee}(\varepsilon)^2 f_{\rm e}(\varepsilon)[1-f_{\rm e}(\varepsilon)] 
  + {\cal I}_{\rm eh}(\varepsilon) {\cal I}_{\rm he}(\varepsilon) [ f_{\rm e}(\varepsilon)(1-f_{\rm h}(\varepsilon)) + f_{\rm h}(\varepsilon)(1-f_{\rm e}(\varepsilon)) ] 
  \right. \nonumber \\ & \left. \mbox{} 
  + {\cal I}_{\rm hh}(\varepsilon)^2 f_{\rm h}(\varepsilon)[1-f_{\rm h}(\varepsilon)] \right\},
\end{align}
where the matrix elements of the current operator ${\cal I}_{\rm ee}(\varepsilon)$, ${\cal I}_{\rm eh}(\varepsilon)$, ${\cal I}_{\rm he}(\varepsilon)$, and ${\cal I}_{\rm hh}(\varepsilon)$ are defined through the matrix relation
\be
  \left( \begin{array}{cc} {\cal I}_{\rm ee}(\varepsilon) & {\cal I}_{\rm eh}(\varepsilon) \\ {\cal I}_{\rm he}(\varepsilon) & {\cal I}_{\rm hh}(\varepsilon) \end{array} \right) =
  \left( \begin{array}{cc} -1 & 0 \\ 0 & 1 \end{array} \right) -
  \left( \begin{array}{cc} r_{\rm ee}(\varepsilon)^* & 
  r_{\rm he}(\varepsilon)^* \\ r_{\rm eh}(\varepsilon)^* &
  r_{\rm hh}(\varepsilon)^* \end{array} \right)
  \left( \begin{array}{cc} -1 & 0 \\ 0 & 1 \end{array} \right) 
  \left( \begin{array}{cc} r_{\rm ee}(\varepsilon) & 
  r_{\rm eh}(\varepsilon) \\ r_{\rm he}(\varepsilon) &
  r_{\rm hh}(\varepsilon) \end{array} \right),
  \label{eq:B2}
\ee
$f_{\rm e}(\varepsilon)$ is the distribution function for the electrons, and $f_{\rm h}(\varepsilon) = 1 - f_{\rm e}(-\varepsilon)$. Combining Eq.\ (\ref{eq:B2}) with unitarity of the Andreev scattering matrix, one finds ${\cal I}_{\rm ee}(\varepsilon) = -2 |r_{\rm he}(\varepsilon)|^2$, ${\cal I}_{\rm eh}(\varepsilon) = {\cal I}_{\rm he}(\varepsilon)^* = - 2 r_{\rm he}(\varepsilon)^* r_{\rm hh}(\varepsilon)$, and ${\cal I}_{\rm hh}(\varepsilon) = 2 |r_{\rm eh}(\varepsilon)|^2$. Omitting contributions proportional to $f_{\rm e}(\varepsilon) (1 - f_{\rm e}(\varepsilon))$ and $f_{\rm h}(\varepsilon) (1 - f_{\rm h}(\varepsilon))$, which vanish at zero temperature, the shot noise power is then found to be
\be
  S = \frac{8 e^2}{h} \int_0^{\infty} d\varepsilon |r_{\rm hh}(\varepsilon)|^2 |r_{\rm he}(\varepsilon)|^2 [ f_{\rm e}(\varepsilon)(1-f_{\rm h}(\varepsilon)) + f_{\rm h}(\varepsilon)(1-f_{\rm e}(\varepsilon)) ].
\ee
Substituting $f_{\rm e}(\varepsilon) = 1$ if $\varepsilon < e V$, with $e V > 0$, and $0$ otherwise, one arrives at the final expression
\be
  S = \frac{8 e^2}{h} \int_0^{e V} d\varepsilon |r_{\rm hh}(\varepsilon)|^2 |r_{\rm he}(\varepsilon)|^2.
\ee

The reflection amplitudes  $r_{\rm hh}(\varepsilon)$ and $r_{\rm eh}(\varepsilon)$ for the one-dimensional Majorana chain have the standard Breit-Wigner form,\cite{Law2009}
\be
  r_{\rm ee}(\varepsilon) = r_{\rm hh}(\varepsilon) = \frac{\varepsilon}{\varepsilon + i \Gamma},\ \
  r_{\rm eh}(\varepsilon) = r_{\rm he}(\varepsilon) = \frac{i \Gamma}{\varepsilon + i \Gamma},
  \label{eq:SBW}
\ee
so that the low-bias shot noise becomes
\be
  S = \frac{8 e^3 |V|}{3 h} \left( \frac{e V}{\Gamma} \right)^2,
\ee
plus corrections of higher order in $eV/\Gamma$. This result has to be multiplied with a factor four to obtain the shot noise power (\ref{eq:Snonint}) of a four-channel Majorana wire.

Similarly, the current of a single non-interacting Majorana chain follows as 
\be
  I = \frac{e}{h} \int_0^{\infty}
  [{\cal I}_{\rm ee} f_{\rm e}(\varepsilon) + {\cal I}_{\rm hh} f_{\rm h}(\varepsilon)],
\ee
which in the zero-temperature limit and for bias $e V > 0$ simplifies to
\be
  I = \frac{2 e}{h} \int_0^{e V} d\varepsilon |r_{\rm he}(\varepsilon)|^2.
\ee
Substituting Eq.\ (\ref{eq:SBW}) for $r_{\rm he}$, one then finds
\be
  I= \frac{2e^2}{h}V\left[1-\frac{1}{3}\left(\frac{eV}{\Gamma}\right)^2\right],
\ee
up to corrections of higher order in $eV/\Gamma$. Multiplication by four gives Eq.\ (\ref{eq:Inonint}).

\end{widetext}

\end{document}